\input epsf
\documentstyle[aps]{revtex}
\tighten
\begin{document}


\preprint{HUTP-02/A037,hep-th/0208094}

\title{Inhomogeneous tachyon decay, light-cone structure and D-brane
network problem in tachyon cosmology} 
\author{Shinji Mukohyama}
\address{
Department of Physics, Harvard University\\
Cambridge, MA, 02138, USA
}
\date{\today}

\maketitle

\begin{abstract} 
 We investigate light-cone structure on the world-volume of an unstable
 D-brane with a tachyon decaying inhomogeneously by using a field
 theoretical description. It is shown that (i) light-cones governing
 open strings are narrower than those governing closed strings and will
 eventually collapse inward in all directions except at kinks, where the
 tachyon remains at the top of its potential; and that (ii) light-cones
 governing open strings at a kink will be narrowed only in the direction
 perpendicular to the kink surface. It is also shown that (iii)
 future-directed light-cones governing open strings near a kink are
 tilted towards the kink, compared with those governing closed
 strings. The result (i) implies that open strings except at kinks are
 redshifted, compared with closed strings, and will eventually cease to 
 be dynamical. On the other hand, the result (ii) shows that open 
 strings on a kink surface can move freely along the kink surface and
 are dynamical but do not feel the existence of the spatial dimension
 perpendicular to the kink surface. The result (iii) indicates that open
 strings near a kink have tendency to move towards the kink. Hence, the
 light-cone structure vividly illustrates how open strings behave during
 the dynamical formation of a kink. We also discuss about a possibility
 that the early universe has a network of various dimensional D-branes,
 black-branes and tachyon matter. A problem associated with the network
 and a possible solution to the problem are discussed.
\end{abstract}


\section{Introduction}


Time-dependent backgrounds are recently getting more important than ever
in the study of string theory. Actually, spacelike
branes~\cite{Gutperle-Strominger} or a rolling
tachyon~\cite{Rolling-tachyon} have been attracting a great deal of
interest. Some of non-BPS objects such as a D-brane-anti-D-brane pair
and a non-BPS D-brane are by their nature unstable and should become  
time-dependent under perturbations. Since non-BPS objects are expected
to play important roles in exploration of string dualities beyond the
BPS level, one would like to study their properties including their
dynamics. Hence, it is important to study the dynamics of tachyons
living on those objects, as their unstable nature is characterized by
the existence of tachyons. 


It is not only in exploration of the string duality web but also in 
exploration of the early universe where tachyons can play important
roles. The so called tachyon cosmology was recently initiated and
investigated by many authors~\cite{tachyon-cosmology}. In this context a
tachyon can dominate our universe at an early epoch and, thus, the
dynamics of tachyons is essential. 


The dynamics of a tachyon can actually be described in many different
ways. For example, it was shown by boundary states of an unstable
D-brane that the pressure of tachyon fluid approaches zero as the tachyon
rolls down~\cite{tachyon-matter}, it is a field theoretical
description~\cite{Garousi,BRWEP,Kluson,tachyon-matter} that has been
extensively used in the tachyon cosmology, and it was by using the
boundary string field theory (BSFT) that some exact properties of the
tachyon condensation were derived in 
refs.~\cite{Garasimov-Shatashvili,Kutasov-Marino-Moore}. 


One of well-known properties of the tachyon effective action for an 
unstable D-brane is that the pressure at late time falls off
exponentially as the tachyon field evolves from any spatially
homogeneous initial configuration towards the minimum of the
potential. This was shown by the boundary states~\cite{tachyon-matter},
in the field theoretical description with a specific form of the tachyon 
potential~\cite{field-theory}, and in the
BSFT~\cite{Sugimoto-Terashima,Minahan} although there are some
differences in details. 


Another well-known property is the absence of plane-wave solutions
around the minimum of the tachyon effective potential.  This is easily
inferred from evidences showing that the minimum describes a 
configuration without D-branes~\cite{no-branes}. The absence of
plane-wave solutions was shown in the field theoretical
description~\cite{field-theory} and the BSFT~\cite{Ishida-Uehara}. 


Knowing the absence of plane-wave solutions around the minimum, it is 
interesting to ask how this is achieved from the open string points of 
view. Before the tachyon decay, open strings are dynamical on the
world-volume of the unstable non-BPS brane. On the other hand, after
the tachyon decay, there are no open string states. Hence, it seems
natural to ask ``what really happens to open strings in the process of
the tachyon decay?'' In this respect, a non-perturbative confinement
mechanism on the brane~\cite{Bergman-Hori-Yi} and the classical
equivalence between a gauge system and a string
fluid~\cite{Gibbons-Hori-Yi} are very suggestive. 


In ref.~\cite{Sen-kink} Sen conjectured that a tachyonic kink on the
world-volume of an unstable D$(p+1)$-brane is a BPS D$p$-brane. In this
context, the tachyon field for the kink configuration is usually
supposed to be static. On the other hand, in the context of the rolling
tachyon or the tachyon cosmology, the tachyon is usually spatially
homogeneous or nearly homogeneous. Hence, it is interesting to
investigate a time-dependent kink, or a highly inhomogeneous rolling
tachyon. It may describe the dynamical formation of the kink as a BPS
D-brane and, thus, may provide us with more knowledge about the dynamics
of the tachyon. Far from the kink, the tachyon condensation is expected
to proceed without being affected by the existence of the kink. Hence,
to this region the previous works on the homogeneous rolling tachyon 
can perhaps be applied. However, it is not clear what dynamically
happens to open strings near a kink.


In this paper we give yet another view on the fate of open strings 
during tachyon decay. This view can be applied to the formation process
of a tachyonic kink, or inhomogeneous tachyon decay. Our strategy in
this paper is to analyze the light-cone structure on the world-volume of
an unstable D-brane by using a field theoretical description. In
particular, we compare light-cones governing open strings with those
governing closed strings. In this view, it is shown that open strings
except at tachyonic kinks are redshifted away. On the other hand, open
strings on a kink surface remain dynamical but do not feel the existence
of the spatial dimension perpendicular to the kink surface. Moreover,
open strings near a kink have tendency to move towards the kink during
the decay process.

We finally discuss in the context of the tachyon cosmology about a
possibility that the early universe has a network of various dimensional
D-branes, black-branes and tachyon matter. A problem associated with the
network and a possible solution to the problem are discussed.


This paper is organized as follows. In Sec.~\ref{sec:action} we review
an effective field theoretical description of an unstable D-brane and
introduce a metric governing open strings. In
Sec.~\ref{sec:tachyon-decay} we analyze an inhomogeneous tachyon decay,
or a time-dependent kink, and the light-cone structure on the
world-volume. In Sec.~\ref{sec:cosmology} we discuss some cosmological
implications including a problem associated with a D-brane network in
the early universe and a possible solution of it. Sec.~\ref{eqn:summary}
is devoted to a summary of this paper.


\section{Tachyon action and two metrics}
\label{sec:action}

Let us consider the following action of an unstable D($p+1$)-brane. 
%
\begin{equation}
 I = -\int d^{p+2}x V(T)
  \sqrt{|\det{\left[g_{MN} + \partial_M T \partial_N T 
  +2\pi l_s^2(F_{MN}+\partial_M\phi\partial_N\phi)\right]}|},
  \label{eng:action}
\end{equation}
where $g_{MN}$ is the induced metric on the world-volume of the brane,
$T$ is the tachyon representing the unstable nature, $F_{MN}$ and $\phi$ 
represent a gauge field and scalar fields confined on the brane,
respectively, and $V(T)$ is the tachyon potential with vanishing minima 
at $T=\pm\infty$.

Note that closed string degrees freedom (such as gravitons and RR fields)
in the bulk couple to $g_{MN}$ directly but that open string degrees of
freedom on the brane ($F_{MN}$ and $\phi$) always couple to $g_{MN}$
through the combination 
%
\begin{equation}
 G_{MN} = g_{MN} + \partial_M T \partial_N T. 
\end{equation}
Hence, we have two different metrics on the brane: $g_{MN}$ governing
closed string degrees of freedom and $G_{MN}$ governing open string
degrees of freedom~\cite{Gibbons-Herdeiro,Mehen-Wecht}. In particular, as
we shall see explicitly, light-cone structures of the two metrics can be 
very different. In the remaining of this paper, we shall set $F_{MN}$
and $\phi$ to be zero. However, we shall keep in mind the fact that
these open string fields on the brane couple to $G_{MN}$.

In the rest of this paper, we set $F_{MN}=0$ and $\phi=0$ in the action
(\ref{eng:action}): 
%
\begin{equation}
 I = -\int d^{p+2}x V(T)\sqrt{|\det{G_{MN}}|}.
\end{equation}
This is the action considered by many authors in the context of tachyon
cosmology. The equation of motion is 
%
\begin{equation}
 \nabla^2T
  -\frac{1}{2}\frac{\partial^MT\partial_M(\partial^NT\partial_NT)}
  {1+\partial^MT\partial_MT} -\frac{V_{,T}}{V} = 0,
\end{equation}
and the stress energy tensor is
%
\begin{equation}
 T_{MN} = -g_{MN}V(T)\sqrt{1+\partial^LT\partial_LT}
  +\frac{V(T)\partial_MT\partial_NT}
   {\sqrt{1+\partial^LT\partial_LT}}. 
   \label{eqn:TMN}
\end{equation}
The stress energy tensor characterizes how the tachyon couples to
gravity. For example, if we simply add the Einstein-Hilbert term to 
the world-volume action or if we consider a spacetime filling brane then
$T_{MN}$ appears in the right hand side of Einstein equation. On the
other hand, if we introduce the Einstein-Hilbert action to a
higher-dimensional bulk then gravity is described by the brane-world
scenario. In any cases, $T_{MN}$ acts as a source of gravity.

In this paper, we investigate general feature of light-cone structure on
the world-volume of the brane during inhomogeneous tachyon decay without
specifying how $T_{MN}$ couples to gravity.


\section{Inhomogeneous tachyon decay}
\label{sec:tachyon-decay}

We would like to investigate light-cone structure on the world-volume of
an unstable D($p+1$)-brane with a tachyon decaying inhomogeneously. For 
simplicity, we consider inhomogeneity in only one spatial
dimension. Namely, we assume $p$-dimensional plane symmetry. A general 
form of the tachyon field and the induced metric within this symmetry is
given by
%
\begin{eqnarray}
 T & = & T(t,y), \nonumber\\
 g_{MN}dx^Mdx^N & = & -N^2(t,y)dt^2 + A^2(t,y)\sum_{i=1}^p(dx^i)^2 + dy^2. 
\end{eqnarray}
For this ansatz, the metric $G_{MN}$ governing open strings is
%
\begin{equation}
 G_{MN}dx^Md^N = -\tilde{N}^2dt^2  
  + A^2\sum_{i=1}^p(dx^i)^2
  + B^2\left(dy-\omega dt\right)^2,
\end{equation}
where
%
\begin{eqnarray}
 \tilde{N}^2 & = & \frac{1-(\dot{T}/N)^2+(T')^2}{1+(T')^2}N^2,
  \nonumber\\
 B^2 & = & 1+(T')^2, \nonumber\\
 \omega & = & -\frac{\dot{T}T'}{1+(T')^2}.
  \label{eqn:tildeN-B-omega}
\end{eqnarray}
Here a dot and a prime denote derivative with respect to $t$ and $y$,
respectively. Note that $\tilde{N}^2\leq N^2$ (the equality holds if and
only if $\dot{T}=0$). 

The future directed null vectors in the $\pm y$ direction are
%
\begin{equation}
 u_{\pm}^M = \left(\frac{\partial}{\partial t}\right)^M 
  \pm N\left(\frac{\partial}{\partial y}\right)^M 
\end{equation}
for $g_{MN}$, and
%
\begin{equation}
 U_{\pm}^M = 
  \left(\frac{\partial}{\partial t}\right)^M 
  +\left(\omega\pm\frac{\tilde{N}}{B}\right)
  \left(\frac{\partial}{\partial y}\right)^M 
  \label{eqn:UpmM}
\end{equation}
for $G_{MN}$. On the other hand, the future
directed null vectors in the $\pm x^i$ direction are 
%
\begin{equation}
 v_{(i)\pm}^M = \left(\frac{\partial}{\partial t}\right)^M 
  \pm \frac{N}{A}\left(\frac{\partial}{\partial x^i}\right)^M 
\end{equation}
for $g_{MN}$, and
%
\begin{equation}
 V_{(i)\pm}^M = \left(\frac{\partial}{\partial t}\right)^M 
  \pm \frac{\tilde{N}}{A}\left(\frac{\partial}{\partial x^i}\right)^M 
\end{equation}
for $G_{MN}$.

From the above expressions of null vectors, we can understand many
things.

(i) First, if $\dot{T}\ne 0$ then the light cones of $G_{MN}$ governing
open strings are narrower in all directions than those of $g_{MN}$
governing closed strings since $\tilde{N}^2<N^2$. In general we expect 
$\dot{T}\ne 0$ except at kinks if we consider the inhomogeneous tachyon
decay. Therefore, open strings except at kinks are redshifted, compared
with closed strings. Moreover, if $\tilde{N}\to 0$ (or $N\to 0$) as
$t\to\infty$ with $y$ fixed, then the light cones of $G_{MN}$ governing
open strings (or those of $g_{MN}$ governing closed strings,
respectively) eventually collapse inward in all directions. When $N\ne
0$, $\tilde{N}$ vanishes if and only if $1-(\dot{T}/N)^2+(T')^2 =
0$. Hence, if 
%
\begin{equation}
 1-(\dot{T}/N)^2+(T')^2\to 0 \quad (t\to\infty, y \mbox{ fixed}), 
  \label{eqn:asymptotic-condition}
\end{equation}
then light-cones governing open strings eventually collapse inward in
all directions while those governing closed strings remain
well-defined. The asymptotic behavior (\ref{eqn:asymptotic-condition}) 
is actually satisfied for a homogeneous rolling tachyon ($|\dot{T}|\to
1$, $T'=0$ and $N=1$). Even for an inhomogeneous tachyon decay, once the
tachyon starts rolling down, the asymptotic behavior
(\ref{eqn:asymptotic-condition}) is generally expected and was
conjectured to be true in ref.~\cite{FKS} based on numerical study of
the inhomogeneous tachyon decay. On the other hand, it is evident that
this condition cannot be satisfied at a kink, where $\dot{T}=0$. Hence,
we can expect the asymptotic behavior (\ref{eqn:asymptotic-condition})
to be true except in the vicinity of kinks. Therefore, open strings
except in the vicinity of kinks are redshifted, compared with closed
strings, and will eventually cease to be dynamical.

(ii) Secondly, at a kink, $v_{(i)\pm}^M=V_{(i)\pm}^M$ since
$\dot{T}=0$. Hence, light-cones of $G_{MN}$ governing open strings at a
kink will be narrowed only in the direction perpendicular to the kink
surface. This, combined with the above consideration (i), implies that
open strings on a kink surface can move freely along the kink surface
and are dynamical but do not feel the existence of the spatial dimension
perpendicular to the kink surface.

\begin{figure}
 \centering\leavevmode\epsfysize=8cm \epsfbox{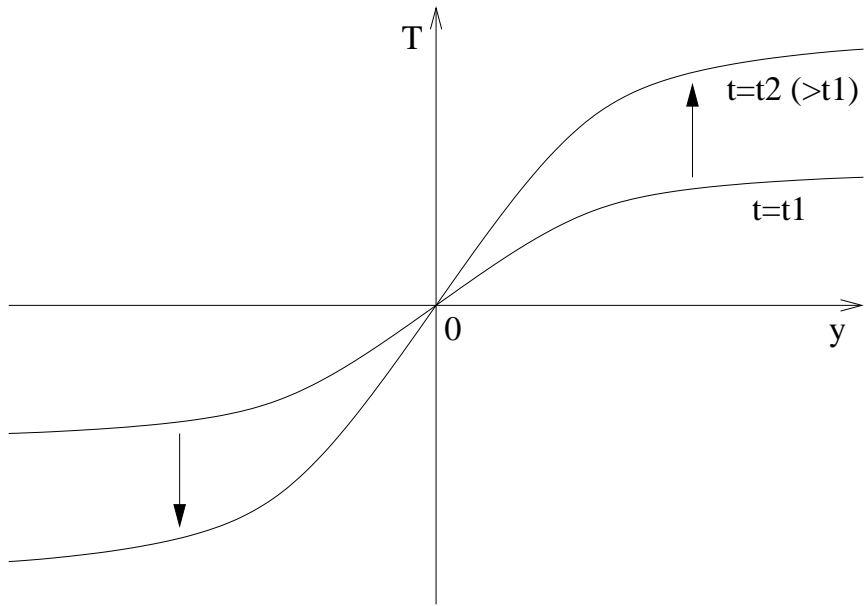}
 \caption{\label{fig:kink}
Time-dependent tachyonic kink.
} 
\end{figure}

(iii) Thirdly, near a kink but not on the kink surface we expect that
$\dot{T}T>0$, provided that the top of the tachyon potential is at
$T=0$. To see this, let us consider a simple example of inhomogeneous
tachyon decay with a sufficiently smooth  initial condition at $t=t_1$
satisfying 
%
\begin{eqnarray}
 T(t_1,y) & = & -T(t_1,-y), \nonumber\\
 \dot{T}(t_1,y) & =& 0,
\end{eqnarray}
and 
%
\begin{eqnarray}
 N(t_1,y) & = & N(t_1,-y), \quad
  \dot{N}(t_1,y) = \dot{N}(t_1,-y), \nonumber\\
 A(t_1,y) & = & A(t_1,-y), \quad
  \dot{A}(t_1,y) = \dot{A}(t_1,-y). 
\end{eqnarray}
Of course, appropriate constraint equations must be imposed on the
initial condition. For the initial condition, we expect the tachyon to
start rolling down the potential hill except at the kink where $T$ and
$\dot{T}$ remain to vanish (see Fig.~\ref{fig:kink}). For this simple
example, it is evident that $\dot{T}T>0$ near the kink but not on the
kink surface. We expect that this statement is true for a wide class of
inhomogeneous tachyon decay if applied to a sufficiently close vicinity
of a kink. Hence, $(y-y_0)\omega<0$ near a kink but not on the kink
surface, where $y_0$ is the position of the kink and $\omega$ is defined
by (\ref{eqn:tildeN-B-omega}). Note that $\omega$ appears in 
(\ref{eqn:UpmM}) and, thus, the null vectors $U_{\pm}^M$ for $G_{MN}$
near a kink are tilted towards the kink, compared with the null vectors
$v_{\pm}^M$ for $g_{MN}$. In other words, future-directed light-cones
governing open strings near a kink are tilted towards the kink, compared
with those governing closed strings. This indicates that open strings
near a kink have tendency to move towards the kink.

In the above three considerations, the light-cone structure vividly
illustrates how open strings behave during the dynamical formation of a
kink.

In ref.~\cite{Sen-kink} Sen conjectured that a tachyonic kink on the 
world-volume of an unstable D$(p+1)$-brane is a BPS D$p$-brane. This
conjecture seems consistent with the time-dependent kink, or the
inhomogeneous tachyon decay, considered above rather than a static
kink. Actually, if we consider a static configuration ($\dot{T}=0$),
$\tilde{N}=N$ and $\omega=0$. Hence, it is easily seen that for a static
configuration, light-cones of $G_{MN}$ governing open strings are
well-defined as far as those of $g_{MN}$ governing closed strings are
well-defined and that light-cones of $G_{MN}$ near a kink are not tilted
towards the kink at all. Hence, open strings on the static kink
background can be as dynamical as closed strings: there is no sign of 
redshift or tendency to move towards the kink. On the other hand, for
the time-dependent kink considered above, all three considerations (i) -
(iii) seem to indicate that a lower-dimensional D-brane is dynamically
formed at the position of the kink.


\section{D-brane network in the early universe}
\label{sec:cosmology}

In the context of the tachyon cosmology, if the universe starts from the
top of the tachyon potential then we expect dynamical formation of the
time-dependent tachyonic kink. In fact, quantum fluctuations kick 
the background tachyon field to different directions at different points
and, as a result, the background tachyon field falls into different
vacua in different regions. In this way, the formation of a network of
kinks is expected. This is exactly like the formation of a network of 
topological defects in the early universe.

Sen's conjecture implies that the kink network is actually a D-brane
network. Hence, if the tachyon cosmology is responsible for an early
stage of the history of our ($4$-dimensional or higher dimensional)
universe, then our universe experienced an era in which there is a
network of D-branes. On one side of a D-brane and another side, the
tachyon is rolling towards its (different) vacua and, hence, there is
what is called tachyon matter in each region.

This is not the end of the story. After the D-brane network formation, a
loop of D-brane can collapse in principle. If this happens then it would
imply a formation of a network of lower-dimensional D-branes since this
process is locally equivalent to a D-brane anti-D-brane
annihilation. Actually, a vortex on the world volume of a coincident 
D$p$-brane and anti D$p$-brane pair is believed to be a
D($p-2$)-brane~\cite{Sen-vortex,Witten}. When the lower dimensional
D-branes are produced, tachyon matter is again expected to be produced
among the newly created D-brane network.

Therefore, we expect that at its early stage, the ($4$-dimensional or
higher dimensional) tachyon cosmology has a very rich structure composed
of networks of various dimensional D-branes and tachyon
matter. Moreover, if the string coupling is strong enough, then these
D-branes can be black-branes. In this case, the D-brane network should
be replaced by a network of black-branes. If the string coupling, or the 
dilation, is spacetime dependent then the network may be composed of
mixture of not only various dimensional D-branes (without horizons) and
tachyon matter but also various dimensional black-branes.

In ref.~\cite{Kofman-Linde} it was argued that the formation of
tachyonic kinks is hazardous in $4$-dimensional cosmology. As mentioned
in the above, the tachyonic kinks are D-branes (or black-branes). Hence,
we can quantitatively investigate how hazardous the formation of the
network of tachyonic kinks in the early universe is.

In the $4$-dimensional standard cosmology, there is a very significant 
constraint on the mass per unit area, or the tension, in a network of
domain walls~\cite{ZKO}. In general a domain wall can annihilate with an
anti-domain wall if the spacetime is flat and we can wait for a
sufficiently long time. However, the spacetime representing the early
universe is not flat but curved. The universe is actually expanding and,
because of the expansion, the annihilation process in the universe
cannot be completed but has to leave on the order of one domain wall
stretching across each Hubble radius. Hence, the mass of the part of the
domain wall network within the present Hubble volume is
$M_{DW}\sim\sigma H_0^{-2}$, where $\sigma$ is the tension of the wall
and $H_0$ is the present value of the Hubble parameter. This mass must
be much smaller than the total mass within the present Hubble volume,
$M_{tot}\sim m_{pl}^2 H_0^{-1}$, since the strongly anisotropic mass
distribution of a domain wall can contribute to the cosmic microwave
background (CMB) anisotropy. Actually, the ratio $M_{DW}/M_{tot}$ must
be smaller than the CMB anisotropy:
%
\begin{equation}
 \frac{M_{DW}}{M_{tot}}\leq 10^{-5}. 
\end{equation}
Hence, we obtain the constraint
%
\begin{equation}
 \frac{\sigma}{m_{pl}^2H_0} \leq 10^{-5}. 
  \label{eqn:constraint}
\end{equation}

Now let us apply the constraint (\ref{eqn:constraint}) to the D-brane
network. For simplicity, we consider an unstable D$9$-brane in
$10$-dimension and, hence, the network of BPS D$8$- (and
anti-D$8$-) branes. We compactify six extra dimensions to obtain
$4$-dimensional universe with a network of $2$-branes. In this case,
$\sigma\sim m_s^9V_6/g_s$ and $m_{pl}^2\sim m_s^8V_6/g_s^2$, where $m_s$ 
is the string scale, $g_s$ is the string coupling constant and $V_6$ is
the volume of the six extra dimensions. Hence, the constraint
(\ref{eqn:constraint}) is reduced to 
%
\begin{equation}
 \frac{g_s^2m_{pl}}{H_0v_6^{1/2}} \leq 10^{-5},
\end{equation}
where $v_6\equiv m_s^6V_6$ is the volume of the six extra dimensions in
the string unit. We have another constraint from the fact that we do not
observe the tower of Kaluza-Klein modes: $m_{KK}/m_{pl}\geq 10^{-16}$,
where $m_{KK}\sim V_6^{-1/6}$ is the mass of the lightest Kaluza-Klein 
mode. This constraint is written as
%
\begin{equation}
 \frac{g_s}{v_6^{2/3}} \geq 10^{-16}. 
\end{equation}
Combining this with the previous constraint, we obtain
%
\begin{equation}
 10^{-32}v_6^{4/3}\leq g_s^2\leq 
  10^{-5}\cdot\frac{H_0}{m_{pl}}\cdot v_6^{1/2}.
\end{equation}
There are values of the string coupling constant $g_s$ allowed by this 
inequality if and only if 
$v_6^{5/6}\leq 10^{27}H_0/m_{pl}\sim 10^{-34}$. This would require that
the extra dimensions should be much smaller than the string
length. Hence, the constraints cannot be satisfied by parameter values
for which the low energy description is valid.

So far we have not yet taken into account effects of RR fields. However,
since an RR field couples to the tachyon through the interaction of the
form 
%
\begin{equation}
 \int f(T)dT\wedge C_{p+1},
\end{equation}
where $f(T)$ is a function of the tachyon field $T$ and $C_{p+1}$ is the
($p+1$)-form potential, the RR field can be excited by the tachyon
decay. Hence, the effect of the RR field is one of important ingredients
of the tachyon cosmology. Progress in this direction and its
cosmological implications will be reported elsewhere~\cite{MST}.

Now we would like to propose a possible solution to the D-brane network
problem addressed in the above. A basic idea is that an RR field can
play the role of a positive cosmological constant.

First, for simplicity, let us consider a spacetime filling brane
($p+2=D$). In this case, after the tachyon decay, the ($p+2$)-form RR
flux is constant and plays the role of a positive cosmological constant 
$\Lambda_{RR}$. Since the RR flux can be excited by the tachyon decay,
we can expect a positive $\Lambda_{RR}$. This may solve the D-brane
network problem in the tachyon cosmology. However, a problem now is how
to reduce $\Lambda_{RR}$ to almost vanishing value after the D-brane 
network problem is solved. We have to rely on other mechanisms like 
Brown-Teitelboim mechanism~\cite{Brown-Teitelboim} to reduce the value
of $\Lambda_{RR}$.

Second, if we consider an unstable brane with less dimensions ($p+2<D$), 
then the ($p+2$)-form after the tachyon decay is no longer a
constant. However, it may still play the role of a positive cosmological
constant for a while if the decay of the RR field is not too rapid. A
question now is whether the time scale of the decay of the RR field is
long enough to solve the D-brane network problem.

In both cases, further investigation of the D-brane network problem and
its solution is necessary~\cite{MST}.

\section{Summary}
\label{eqn:summary}

We have investigated light-cone structure on the world-volume of an
unstable D-brane with a tachyon decaying inhomogeneously by using a field 
theoretical description. It has been shown that (i) light-cones
governing open strings are narrower than those governing closed strings
and will eventually collapse inward in all directions except at kinks,
where the tachyon remains at the top of its potential; that (ii)
light-cones governing open strings at a kink will be narrowed only in
the direction perpendicular to the kink surface; and that (iii)
future-directed light-cones governing open strings near a kink are
tilted towards the kink, compared with those governing closed
strings. The result (i) implies that open strings except at kinks are 
redshifted, compared with closed strings, and will eventually cease to 
be dynamical. On the other hand, the result (ii) shows that open 
strings on a kink surface can move freely along the kink surface and
are dynamical but do not feel the existence of the spatial dimension
perpendicular to the kink surface. The result (iii) indicates that open
strings near a kink have tendency to move towards the kink. In this way,
the light-cone structure vividly illustrates how open strings behave
during the dynamical formation of a kink.

Finally, we have also discussed about a possibility that the early
universe has a network of various dimensional D-branes, black-branes and
tachyon matter. An associated problem was pointed out and a possible
solution of the problem was proposed.

\begin{acknowledgments}
 This work was initiated during a visit to DAMTP, Cambridge
 University. The author would like to thank DAMTP for hospitality and
 Gary Gibbons for stimulating discussions. He is grateful to Lev Kofman, 
 Kazutoshi Ohta and Shigeki Sugimoto for their helpful comments and
 discussions. He is thankful to Werner Israel for his continuing
 encouragement. This work is supported by JSPS Postdoctoral Fellowship
 for Research Abroad. 
\end{acknowledgments}


\end{document}